\documentclass[preprintnumbers,article,amsmath,amssymb,floatfix,10pt,prd,twocolumn,
superscriptaddress,nofootinbib]{revtex4-2}
\usepackage{bm}
\usepackage{amsfonts}
\usepackage{latexsym}
\usepackage[latin1]{inputenc}
\usepackage{graphicx}
\usepackage{amsmath}
\usepackage{palatino}
\usepackage{mathpazo}
\usepackage{textcomp}
\usepackage{float}
\usepackage{booktabs}
\usepackage{dcolumn}
\usepackage{ragged2e}
\usepackage{hyperref}
\hypersetup{colorlinks,citecolor=blue}
\hypersetup{colorlinks=true,linkcolor=red,filecolor=magenta,    urlcolor=blue}
\usepackage{amsmath}
\usepackage{xcolor}
\usepackage{orcidlink}
\usepackage{epsfig}
\usepackage{caption}
\usepackage{subcaption}
\usepackage{commath}
\def\jnl@style{\it}
\def\aaref@jnl#1{{\jnl@style#1}}

\def\aaref@jnl#1{{\jnl@style#1}}

\def\aj{\aaref@jnl{AJ}}                   
\def\apj{\aaref@jnl{ApJ}}                 
\def\apjl{\aaref@jnl{ApJ}}                
\def\apjs{\aaref@jnl{ApJS}}               
\def\apss{\aaref@jnl{Ap\&SS}}             
\def\aap{\aaref@jnl{A\&A}}                
\def\aapr{\aaref@jnl{A\&A~Rev.}}          
\def\aaps{\aaref@jnl{A\&AS}}              
\def\mnras{\aaref@jnl{Mon.~Not.~Roy.~Astron.~Soc.}}             
\def\prd{\aaref@jnl{Phys.~Rev.~D}}        
\def\prc{\aaref@jnl{Phys.~Rev.~C}}  
\def\prl{\aaref@jnl{Phys.~Rev.~Lett.}}    
\def\qjras{\aaref@jnl{QJRAS}}             
\def\skytel{\aaref@jnl{S\&T}}             
\def\ssr{\aaref@jnl{Space~Sci.~Rev.}}     
\def\zap{\aaref@jnl{ZAp}}                 
\def\nat{\aaref@jnl{Nature}}              
\def\aplett{\aaref@jnl{Astrophys.~Lett.}} 
\def\apspr{\aaref@jnl{Astrophys.~Space~Phys.~Res.}} 
\def\physrep{\aaref@jnl{Phys.~Rep.}}      
\def\physscr{\aaref@jnl{Phys.~Scr}}       
\def\commat{\aaref@jnl{Comm.~Math.~Phys.}}              
\def\science{\aaref@jnl{Science}}               
\def\cqg{\aaref@jnl{Classical Quant.~Grav.}}            
\def\jpcs{\aaref@jnl{JPCS}}                                     
\def\ijmpd{\aaref@jnl{Int.~J.~Mod.~Phys.~D}}                    
\def\grg{\aaref@jnl{Gen.~Relat.~Gravit.}}               
\def\rpp{\aaref@jnl{Rep.~Prog.~Phys.}}          
\def\npa{\aaref@jnl{Nucl.~Phys.~A}}        
\def\lrr{\aaref@jnl{Living Rev.~Rel.}}                   
\def\jcap{\aaref@jnl{J.~Cosmology Astropart.~Phys.}}    
\def\rmp{\aaref@jnl{Rev.~Mod.~Phys.}}   
\def\epjc{\aaref@jnl{Eur.~Phys.~J.~C}} 
\def\plb{\aaref@jnl{~Phy.~Lett.~B}} 
\def\mpla{\aaref@jnl{Mod.~Phy.~Lett.~A}} 
\def\arxiv{\aaref@jnl{arxiv.org}}

\allowdisplaybreaks[1]

\begin{document}
\color{black}       
\title{Dark energy constraint on equation of state parameter in the Weyl type $f(Q,T)$ gravity}

\author{Gaurav N. Gadbail\orcidlink{0000-0003-0684-9702}}
\email{gauravgadbail6@gmail.com}
\affiliation{Department of Mathematics, Birla Institute of Technology and
Science-Pilani,\\ Hyderabad Campus, Hyderabad-500078, India.}

\author{Simran Arora\orcidlink{0000-0003-0326-8945}}
\email{dawrasimran27@gmail.com}
\affiliation{Department of Mathematics, Birla Institute of Technology and
Science-Pilani,\\ Hyderabad Campus, Hyderabad-500078, India.}

\author{P.K. Sahoo\orcidlink{0000-0003-2130-8832}}
\email{pksahoo@hyderabad.bits-pilani.ac.in}
\affiliation{Department of Mathematics, Birla Institute of Technology and
Science-Pilani,\\ Hyderabad Campus, Hyderabad-500078, India.}
%

\begin{abstract}
The equation of state parameter is a significant method for characterizing dark energy models. We investigate the evolution of the equation of state parameter with redshift using a Bayesian analysis of recent observational datasets (the Cosmic Chronometer data (CC) and Pantheon samples). The Chevallier-Polarski-Linder parametrization of the effective equation of state parameter, $\omega_{eff}=\omega_0+\omega_a \left( \frac{z}{1+z}\right) $, where $\omega_0$ and $\omega_a$ are free constants, is confined to the Weyl type $f(Q,T)$ gravity, where $Q$ represents the non-metricity and $T$ is the trace of the energy-momentum tensor. We observe the evolution of the deceleration parameter $q$, the density parameter $\rho$, the pressure $p$, and the effective equation of state parameter $\omega$. The cosmic data limit for $\omega$  does not exclude the possibility of $\omega < -1$. It is seen that the parameter $\omega$ shows a transition from deceleration to acceleration, as well as a shift from $\omega>-1$ to $\omega<-1$. \\


\end{abstract}

\date{\today}
\maketitle

\section{Introduction}

Observations evidenced by astronomical probes like type Ia supernovae  \cite{Perlmutter/1999,Riess/1998,Riess/2004}, cosmic microwave background radiation \cite{Komatsu/2011,Huang/2006}, and large-scale structure \cite{Koivisto/2006, Daniel/2008} reveal that the universe transitioned from an early deceleration to a  recent acceleration phase. Finding the responsible candidate for the late-time cosmic accelerated expansion is one of the most sensitive issues in modern cosmology. Our universe is dominated by an unknown form of energy termed as ``dark energy" (DE) \cite{Peebles/1988,Ratra/1988,Steinhardt/1999}. Although the incorporation of DE, such as the cosmological constant, has proven extremely effective, it is hindered by theoretical issues of fine-tuning and cosmic coincidence \cite{Peebles/2003,Sahni/2000}. \\
An alternative approach to dark energy is to modify the gravitational part of the Einstein-Hilbert action, called the modified theory of gravity. If we acknowledge the geometrical character of gravity as argued by the equivalence principle, it is necessary to investigate how gravity can be geometrized in an equivalent manner. When a flat spacetime with metric but asymmetric connections is considered, an equivalent description of general relativity (GR) emerges. This work aims at the symmetric teleparallel representation of GR, which is built on an equally flat spacetime and attributes gravity to non-metricity $Q$ \cite{Nester/1999}. In the context of proper Weyl geometry, we consider the extension of the $f(Q)$ gravity \cite{Jimenez/2018,Lazkoz/2019,Atayde/2021,Mandal/2020}, i.e. the $f(Q,T)$ gravity \cite{Yixin/2019a,Najera/2021, Arora/2021,Najeraa/2022,Gadbail/2022a}, where $Q$ is the nonmetricity and $T$ is the trace of energy-momentum tensor. The cosmic implications of the Weyl type $f(Q,T)$ gravity have been analyzed and considered as an alternative for describing cosmological early and late phases of evolution \cite{Yixin/2020b,Yang/2021, Gadbail/2021a,Gadbail/2021b}. In the framework of the proper Weyl geometry, the scalar non-metricity $Q$ is wholly determined by the magnitude of the Weyl vector $w_{\lambda}$. For the flat geometry constraint, the total scalar curvature vanishes in the Weyl geometry and adds this condition to the gravitational action via a Lagrangian multiplier, $\lambda$. Recently, the Weyl gravity has seen a resurgence in order to solve the dark energy, dark matter issues, and the inflation mentioned in \cite{Alvarez/2017}.\\
Although multiple observations have confirmed the presence of DE, its nature remains a mystery to us. The condition to accelerate the expansion is $\omega<-\frac{1}{3}$, as defined by the equation of state parameter. To understand the gravity or dynamics of the universe, the physics underpinning DE determines the equation of state \cite{Carroll/2003,Gong/2007,Huang/2008}. As a result, this work brings together the parametrized EoS and the modified Weyl type $f(Q,T)$ gravity.\\ 
In the literature, one can find many different EoS parametrizations. One of the simplest and earliest parametrizations presented by Chevallier-Polarski-Linder is the so-called CPL parametrization \cite{Chevallier/2001}. The CPL parametrization is the Taylor expansion of $\omega$ with respect to the scale factor $a$ up to the first order as $\omega (a)=\omega_0+\omega_a(1-a)$ and consequently in terms of redshift as $\omega(z)=\omega_0+\omega_a(\frac{z}{1+z})$. Notice that although the CPL is a well-behaved parametrization at early $(z\to \infty)$ and present $ (z = 0)$ epochs, it diverges at future time $(z = -1)$. This parameter behaves well at high redshifts and is a good approximation for DE slow roll scalar field models \cite{Linder/2003,Gong/2007}. As a  result, utilizing observational data to constrain the two-parameter CPL EoS in the Weyl type $f(Q,T)$ gravity is intriguing. We use the Cosmic Chronometer (CC) $Hubble$ data,  the $Pantheon$ samples (SNe Ia) and the $BAO$ data for this purpose. We notice that the current best fit for the EoS ($\omega$) is less than -1, implying phantom dark energy and an increase in energy density with time.\\
The outline is as follows: In section \ref{section 2}, we discuss the Weyl-type $f(Q, T)$ gravity formalism. In section \ref{section 3}, we obtained the expression for the Hubble parameter using the two-parameter equation of state in the Friedmann-Lemaitre-Robertson-Walker (FLRW) framework. In Section \ref{section 4}, we constrain the model parameters using the $Hubble$ data, the $Pantheon$ data, and the combination ($Hz+Pantheon$ and $BAO+Hz+Pantheon$) by the MCMC technique. We analyze the behavior of cosmological parameters in section \ref{section 5}. Lastly, in section \ref{section 6}, the obtained results are discussed.

\section{Overview of Weyl type $f(Q,T)$ Gravity}
\label{section 2}

However, we have adapted to the conventional formulation of GR in which gravity is linked to spacetime curvature. There are two different ways to formalize GR in flat spacetime: torsion or non-metricity. A  formulation of GR in flat, torsionless spacetime is symmetric teleparallel gravity. \\
In 1918, Weyl suggested a novel geometry by proposing a relationship with the feature that under parallel vector transport, both the orientation and magnitude of a vector change \cite{Weyl/1918}. In a  Weyl geometry, the connection is no longer metric compatible. Furthermore, the Weyl connection in terms of a new vector field known as the Weyl vector field is given as

\begin{equation}
\label{Connection}
\widetilde{\Gamma}^\lambda_{\mu\nu}\equiv\Gamma^\lambda_{\mu\nu}+g_{\mu\nu}w^\lambda-\delta^\lambda_\mu w_\nu-\delta^\lambda_\nu w_\mu .
\end{equation}
which results in $\widetilde{\nabla}_{\lambda}g_{\mu \nu}= 2w_{\lambda}g_{\mu\nu}$.\\
The Weyl type $f(Q,T)$ gravity is described by the action \cite{Yixin/2020b}
 \begin{multline}
\label{1}
 S=\int d^4x\sqrt{-g}\left[ \kappa^2f(Q,T)-\frac{1}{4}W_{\mu\nu}W^{\mu\nu}-\frac{1}{2}M^2w_\mu w^\mu+\right.\\ 
 \left. \lambda\left(R+6\nabla_\alpha w^\alpha-6w_\alpha w^\alpha\right)+\mathcal{L}_m\right],
\end{multline}    
with  $\kappa^2=\frac{1}{16\pi G}$. Here, $f(Q,T)$ is an arbitrary function of the nonmetricity, and the trace of the energy-momentum tensor. The particle's mass to the vector field is denoted by $M$, and $g=det(g_{\mu\nu})$. The scalar non-metricity is defined as  

\begin{equation}
\label{2}
Q\equiv- g^{\mu\nu}\left(L^\alpha_{\beta\nu}L^\beta_{\nu\alpha}-L^\alpha_{\beta\alpha}L^\beta_{\mu\nu}\right),
\end{equation}
where $L^\lambda_{\mu\nu}$ is the deformation tensor read as
\begin{equation}
\label{3}
L^\lambda_{\mu\nu}=-\frac{1}{2}g^{\lambda\gamma}\left(Q_{\mu\gamma\nu}+Q_{\nu\gamma\mu}-Q_{\gamma\mu\nu}\right).
\end{equation}
We define the nonmetricity tensor $Q_{\alpha\mu\nu}$ as the covariant derivative of the metric tensor with respect to $\widetilde{\Gamma}_{\mu\nu}^{\lambda}$,
\begin{equation}
\label{4}
Q_{\alpha\mu\nu}\equiv\widetilde{\nabla}_\alpha g_{\mu\nu}=\partial_\alpha g_{\mu\nu}-\widetilde{\Gamma}^\rho_{\alpha\mu}g_{\rho\nu}-\widetilde{\Gamma}^\rho_{\alpha\nu}g_{\rho\mu}=2w_\alpha g_{\mu\nu}.
\end{equation}\\
Plugging Eq. \eqref{3} in Eq. \eqref{2}, we acquire the important relation 
\begin{equation}
\label{5}
Q=-6w^2.
\end{equation}
Further, the generalized proca equation by varying the action \eqref{1} with respect to vector field is 
\begin{equation}
\label{6}
\nabla^\nu W_{\mu\nu}-\left(M^2+12\kappa^2f_Q+12\lambda\right)w_\mu=6\nabla_\mu \lambda.
\end{equation}
Comparing equation \eqref{6} with the standard Proca equation, we obtaine the effective dynamical mass of the vector field as
\begin{equation}
\label{7}
M^2_{eff}=M^2+12\kappa^2f_Q+12\lambda.
\end{equation}

The generalised field equations obtained by varying the action \eqref{1} with respect to the metric tensor are
\begin{multline}
\label{8}
\frac{1}{2}\left(T_{\mu\nu}+S_{\mu\nu}\right)-\kappa^2f_T\left(T_{\mu\nu}+\Theta_{\mu\nu}\right)=-\frac{\kappa^2}{2}g_{\mu\nu}f\\-6k^2f_Q w_\mu w_\nu +
\lambda\left(R_{\mu\nu}-6w_\mu w_\nu +3g_{\mu\nu}\nabla_\rho w^\rho \right)
\\+3g_{\mu\nu}w^\rho \nabla_\rho \lambda 
-6w_{(\mu}\nabla_{\nu )}\lambda+
g_{\mu\nu}\square \lambda-\nabla_\mu\nabla_\nu \lambda,
\end{multline}
in which 
 \begin{equation}
\label{10}
f_T\equiv \frac{\partial f(Q,T)}{\partial T}, \hspace{0.2in}
f_Q\equiv\frac{\partial f(Q,T)}{\partial Q}.
\end{equation}
Also, the definition of $T_{\mu\nu}$ and $\Theta_{\mu\nu}$ is
\begin{equation}
\label{9}
T_{\mu\nu}\equiv-\frac{2}{\sqrt{-g}}\frac{\delta(\sqrt{-g}L_m)}{\delta g^{\mu\nu}},
\end{equation} 

\begin{equation}
\label{11}
\Theta_{\mu\nu}=g^{\alpha\beta}\frac{\delta T_{\alpha\beta}}{\delta g_{\mu\nu}}=g_{\mu\nu}L_m-2T_{\mu\nu}-2g^{\alpha\beta}\frac{\delta^2 L_m}{\delta g^{\mu\nu}\delta g^{\alpha\beta}}.
\end{equation}
Here, $S_{\mu\nu}$ is the re-scaled energy momentum tensor of the free Proca field given by
\begin{equation}
\label{12}
S_{\mu\nu}=-\frac{1}{4}g_{\mu\nu}W_{\rho\sigma}W^{\rho\sigma}+W_{\mu\rho}W^\rho_\nu -\frac{1}{2}M^2g_{\mu\nu}w_\rho w^\rho +M^2 w_\mu w_\nu ,
\end{equation}
and 
\begin{equation}
\label{13}
W_{\mu\nu}=\nabla_\nu w_\mu-\nabla_\mu w_\nu.
\end{equation}

It is also noted that the expression for the divergence of the matter energy-momentum tensor in the Weyl-type $f(Q,T)$ theory is given by \cite{Yixin/2020b}
\begin{equation*}
\nabla^{\mu}T_{\mu\nu}= \frac{\kappa^2}{1+2\kappa^2f_T}\left[2\nabla_{\nu}(f_T\mathcal{L}_m)-f_T\nabla_{\nu}T-2T_{\mu\nu}\nabla^{\mu}f_T\right]   
\end{equation*}
As a result, the above equation shows that the matter energy-momentum tensor is not conserved in the Weyl-type $f(Q,T)$ theory. The non-conservation of the matter energy-momentum tensor can be interpreted physically as indicating the presence of an extra force acting on massive test particles, causing the motion to be non-geodesic. From a physical perspective, it indicates the amount of energy that enters or leaves a specified volume of a physical system. Moreover, the non-vanishing right-hand side of the energy-momentum tensor indicates the transfer processes or particle production in the system.
One can note that the energy-momentum tensor becomes conserved in the absence of $f_{T}$ terms in the above equation \cite{Jimenez/2018}.

\section{The cosmological model} \label{section 3}

Let us consider that the universe is described by homogeneous, isotropic and spatially flat FLRW line element as
\begin{equation}
\label{14}
ds^2=-dt^2+a^2(t)\delta_{ij}dx^i dx^j ,
\end{equation}
where $a(t)$ is the cosmic scale factor.\\
Assuming the vector field $w_\mu$ as $w_\mu=\left[\psi (t),0,0,0\right]$ \cite{Yixin/2020b} implying $w^2=w_\mu w^\mu=-\psi^2(t)$, and  $Q=-6w^2=6\psi^2(t)$.\\
 The energy momentum tensor for the perfect fluid is defined as:
\begin{equation}
\label{16}
T_{\mu\nu}=\left(\rho+p\right)u_\mu u_\nu+ p g_{\mu\nu},
\end{equation}
where $p$ and $\rho$ are the pressure and the matter energy density, respectively. The four velocity vector $u^\mu$ is such that $u_\mu u^\mu=-1$. Thus implies  $T^\mu_\nu=diag\left(-\rho,p,p,p\right)$, and $\Theta^\mu_\nu=\delta^\mu_\nu p-2T^\mu_\nu=diag\left(2\rho+p,-p,-p,-p\right)$. 

 The flat space constraint and the generalized Proca equation in cosmological case can be represented as
\begin{equation}
\label{19}
\dot{\psi}=\dot{H}+2H^2+\psi^2-3H\psi,
\end{equation}
\begin{equation}
\label{20}
\dot{\lambda}=\left(-\frac{1}{6}M^2-2\kappa^2f_Q-2\lambda\right)\psi=-\frac{1}{6}M^2_{eff}\psi ,
\end{equation}
\begin{equation}
\label{21}
\partial_i \lambda=0.
\end{equation}
From equation \eqref{8} and using given metric \eqref{14} the obtained generalized Friedmann equations are,
\begin{multline}
\label{22}
\kappa^2f_T\left(\rho+p\right)+\frac{1}{2}\rho=\frac{\kappa^2}{2}f-\left(6\kappa^2f_Q+\frac{1}{4}M^2\right)\psi^2 \\
-3\lambda\left(\psi^2-H^2\right)-3\dot{\lambda}\left(\psi-H\right),
\end{multline}

\begin{multline}
\label{23}
-\frac{1}{2}p=\frac{\kappa^2}{2}f+\frac{M^2\psi^2}{4}+\lambda\left(3\psi^2+3H^2+2\dot{H}\right)\\
+\left(3\psi+2H\right)\dot{\lambda}+\ddot{\lambda}.
\end{multline}

For our investigation, we consider the functional form $f(Q,T)=\alpha Q+\frac{\beta}{6\kappa^2}T$, where $\alpha$ and $\beta$ are model parameters. This particular functional form of $f(Q,T)$ is motivated, for instance, in reference \cite{Yixin/2020b}. For certain choice of model parameters, this model is basically equivalent to $\Lambda$CDM model for certain redshift range.

Using this form, we rewrite the field equations \eqref{22} and \eqref{23} as
\begin{equation}
\label{24}
-\left(\frac{\beta}{4}+\frac{1}{2}\right)\rho+\frac{\beta}{4}p=3\alpha \kappa^2 \psi^2+\frac{M^2\psi^2}{4}+3\kappa^2\left(\psi^2-H^2\right),
\end{equation}
\begin{equation}
\label{25}
-\left(\frac{\beta}{4}+\frac{1}{2}\right)p+\frac{\beta}{4}\rho=3\alpha \kappa^2 \psi^2+\frac{M^2\psi^2}{4}+\kappa^2\left(3\psi^2+3H^2+2\dot{H}\right).
\end{equation}
Using the relation $\nabla_{\lambda}g_{\mu\nu}=-w_{\lambda}g_{\mu\nu}$ and $w_1=\psi(t)$, we obtained $\psi(t)=-6H(t)$. Further simplifying Eq.\eqref{24} and \eqref{25}, we get

\begin{multline}
\label{26}
p=-\left(36\left(\frac{18}{\beta+3}\left(\alpha+1\right)+\frac{3\tilde{M}^2}{2\left(\beta+3\right)}\right)+\frac{18}{2\beta+3}\right)H^2\\
-\frac{18\left(\beta+2\right)}{\left(2\beta+3\right)\left(\beta+3\right)}\dot{H},
\end{multline}
and 
\begin{multline}
\label{27}
\rho=\left(\frac{-9\left(11\beta+24\right)\left(24\alpha+25\right)}{4\left(\beta+2\right)\left(\beta+3\right)}+\frac{29\beta+72}{2\left(2\beta+3\right)\left(\beta+2\right)}\right)H^2\\
-\frac{9\beta}{2\left(2\beta+3\right)\left(\beta+3\right)}\dot{H}.
\end{multline}
 where $\tilde{M}^2= M^{2}/\kappa^2$, $\tilde{M}$ is the mass of the Weyl vector field, indicating the strengths of the Weyl geometry-matter coupling. In this case, we have assumed $\tilde{M}= 0.95$ \cite{Yixin/2020b}.\\
The effective equation of state $\omega=\frac{p}{\rho}$ becomes
\begin{equation}
\label{28}
\omega=\frac{-a H^2-b\dot{H}}{cH^2-d\dot{H}},
\end{equation}
where the coefficients $a$, $b$, $c$, and $d$ are as follows
\begin{equation}
\label{29}
a=\frac{18}{2 \beta +3}+36 \left(\frac{18 (\alpha +1)}{\beta +3}+\frac{3 \tilde{M}^2}{2 (\beta +3)}\right),
\end{equation}
\begin{equation}
\label{30}
b=\frac{18 (\beta +2)}{(2 \beta +3) (\beta +3)},
\end{equation}
\begin{equation}
\label{31}
c=\frac{29 \beta +72}{2 (2 \beta +3) (\beta +2)}-\frac{9(24 \alpha +25) (11 \beta +24)}{4 (\beta +2) (\beta +3)},
\end{equation}
\begin{equation}
\label{32}
d=\frac{9 \beta }{2 (2 \beta +3) (\beta +3)}.
\end{equation}
The derivative of the Hubble parameter with respect to time can be written in the form of 
\begin{equation}
\label{33}
\dot{H}=\frac{dH}{dt}=-\left(1+z\right)H(z)\frac{dH}{dz}.
\end{equation}

We need one more ansatz to get the solution to $H$. In this work, a parameterization of the effective equation of state is assumed. We consider the widely used Chevallier-Polarski-Linder (CPL) parametric form of equation of state parameter $\omega$ in terms of redshift $z$  \cite{Chevallier/2001,Linder/2003} 
\begin{equation}
\label{34}
\omega(z) =\omega_0+\omega_a\left(\frac{z}{1+z}\right),
\end{equation}
where $\omega_0$ and $\omega_a$ are constants. The CPL parameterization can be thought of as the Taylor series expansion of $\omega$ up to the first order with respect to the scale factor $a$. It can be seen that $(\omega_0,\omega_a)=(-1,0)$, simplifies the effective equation of state to the $\Lambda$CDM model, and it also converges for large redshifts. The CPL parametric form has various advantages, including a manageable two-dimensional space, excellent accuracy in reconstructing numerous scalar field equations of state and the resulting distance-redshift relations,  and high sensitivity to observational data \cite{Linder/2003,Linder/2003b,Hao/2013}.\\
From equation \eqref{28}, \eqref{33}, and \eqref{34}, we have the following differential equation:
\begin{equation}
\label{35}
\frac{dH}{dz}=-\frac{\left(a+\omega_0\,c+\left(\frac{z}{z+1}\right)\omega_a\, c\right)}{\left(1+z\right)\left(\omega_0\,d-b+\left(\frac{z}{z+1}\right)\omega_a\, d\right)}H(z).
\end{equation}
Solving equation \eqref{35} yields the solution
\begin{equation}
\label{36}
H(z)=H_0 (z+1)^{-\frac{c}{d}} \left(\frac{d (\omega_0\, z+\omega_0+\omega_a\, z)-b (z+1)}{d \omega_0-b}\right)^{l},
\end{equation}
where $l=-\frac{a d+b c}{d (d (\omega_0+\omega_a)-b)}$ and $H(0)=H_0$.\\
The deceleration parameter $q$ defined as $q=-1-\frac{\dot{H}}{H^2}$ is obtained as follows
\begin{equation}
\label{37}
q\left(z\right)=-1-\frac{a (1 + z) + c (\omega_0 +\omega_0\,  z + \omega_a \,z)}{-b (1 + z) + d (\omega_0 + \omega_0\, z + \omega_a\, z)}. 
\end{equation}

\section{observational data}
\label{section 4}

In this section, we will go over the cosmological data that was utilized in this investigation. We employ various contemporary observational data to constrain the model parameters in $H(z)$ using the MCMC technique. We focus on data relevant to the expansion history of the universe, i.e., those characterizing distance-redshift relation. We will specifically use the data from the early-type galaxies (direct $Hubble$ parameter measurements) and the type Ia supernovae ($Pantheon$ samples) spanning Supernova Legacy  Survey (SNLS), Sloan Digital Sky Survey (SDSS), Hubble Space Telescope (HST) survey, Panoramic Survey Telescope and Rapid Response System (Pan-STARRS1). In the following context, for simplicity, we denote the model parameters $\omega_0=m$ and $\omega_a=n$.

\subsection{Hubble data}

The Hubble parameter estimates for early-type galaxies with passive evolution have been yielded by predicting their differential evolution. Compilations of such data are known as cosmic chronometers (CC) \cite{Moresco/2015,Moresco/2018}. We employ a sample of CC covering the redshift range $0<z<1.97$. We examine the constraints on model parameters by the $\chi^{2}$ estimator as follows:

\begin{equation}
\chi^{2}_{Hub}= \sum_{i} \frac{\left( H(\theta_{s},z_{i})-H_{obs}(z_{i})\right) ^{2}}{\sigma^{2}_{Hub}(z_{i})}
\end{equation}
where $\sigma^{2}_{Hub}(z_{i})$ is the standard error on the measured values of $H_{obs}(z_{i})$, and $\theta_{s}$ is the cosmological background parameter space.

\subsection{Pantheon data}
The Pantheon compilation \cite{Scolnic/2018} is one of the most recent type Ia supernovae (SNeIa) data compilations. We consider the set of 1048 SNe, which covers the redshift range $0.01<z<2.26$, and define the $\chi^{2}$ as 
\begin{equation}
\chi^{2}_{SN}= \mu_{SN} C^{-1} \mu^{T}_{SN},
\end{equation}
where $\mu_{SN}= \mu_{i}-\mu_{th}(\theta_{s},z_{i})$ and $\mu_{i}=\mu_{B,i}-M$. Here, $\mu_{B,i}$ is the apparent maximum magnitude for redshift $z_{i}$, $M$ is the hyper-parameter that quantifies uncertainties of various origins. It is used instead of free parameters 
$\alpha$, $\beta$ in the perspective of the "BEAMS" with Bias Corrections method \cite{Kessler/2017}. The theoretical distance modulus is given as 
\begin{equation}
\mu_{th}= 5log\left( \frac{d_{L}(\theta_{s},z)}{Mpc}\right) +25, 
\end{equation}
and 
\begin{equation}
d_{L}(\theta_{s},z)= c(1+z)\int_{0}^{z} \frac{dx}{H(\theta_{s},x)}.
\end{equation}

\subsection{BAO data}
We use the collection of $6dFGS$, $SDSS$, and $Wiggle\,Z$ surveys at various redshifts for BAO data. Here, we employ $\frac{d_{A}}{D_{v}}$ and the following cosmology to establish BAO constraints. 
\begin{eqnarray}
d_{A}(z) &=& c \int_{0}^{z} \frac{dz'}{H(z')},\\
D_{v}(z) &=& \left[\frac{d_{A}^2 \, c\, z}{ H(z)}   \right]^{1/3},\\
\chi^{2} &=& X^{T} C_{BAO}^{-1} X.
\end{eqnarray}
Here, $d_{A}(z)$ represents the comoving angular diameter distance, and $D_{v}$ is the dilation scale. $X$ depends on the survey considered and $C_{BAO}$ is the covariance matrix \cite{Giostri/2012}.

We plot the contours of the $1-\sigma$ and $2-\sigma$ confidence levels from the Hubble and Pantheon data in figure \ref{Combine}. Also, the constraints from the combination $Hz+Pantheon$ and $BAO+Hz+Pantheon$ are given in figure \ref{Hz+Pan} and \ref{BAO+Hz+Pan} by minimizing $\chi^{2}_{Hub}+\chi^{2}_{SN}$ and $\chi^{2}_{BAO}+\chi^{2}_{Hub}+\chi^{2}_{SN}$, respectively. The error bar for the Hubble parameter with the standard $\Lambda$CDM model is shown in figure \ref{error}.

\begin{widetext}

\begin{figure}[H]
\centering
\includegraphics[scale=0.6]{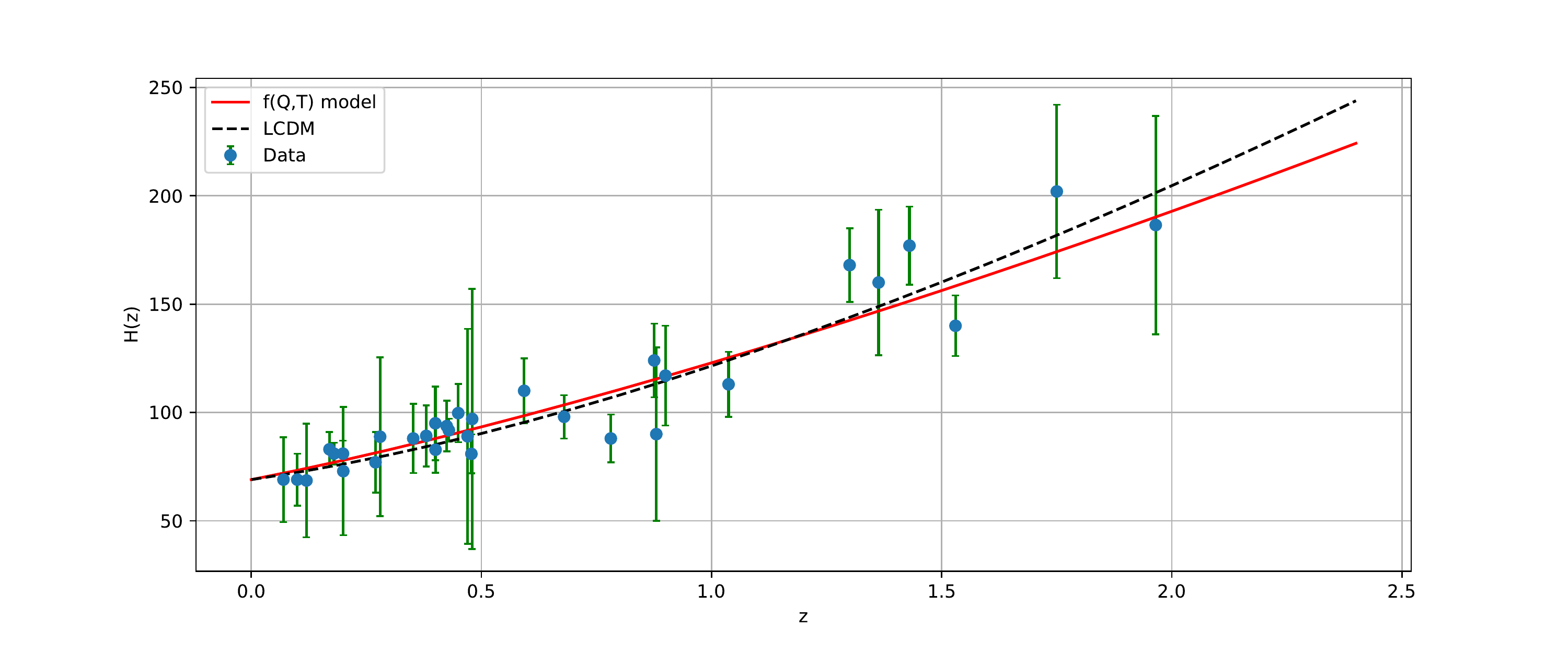}
\caption{The evolution of Hubble parameter with respect to redshift $z$. The blue dots represent error bars, the red line is the curve obtained for our model while the black dashed line corresponds to $\Lambda$CDM model.}
\label{error}
\end{figure}

\begin{figure}[H]
\centering
\includegraphics[scale=0.6]{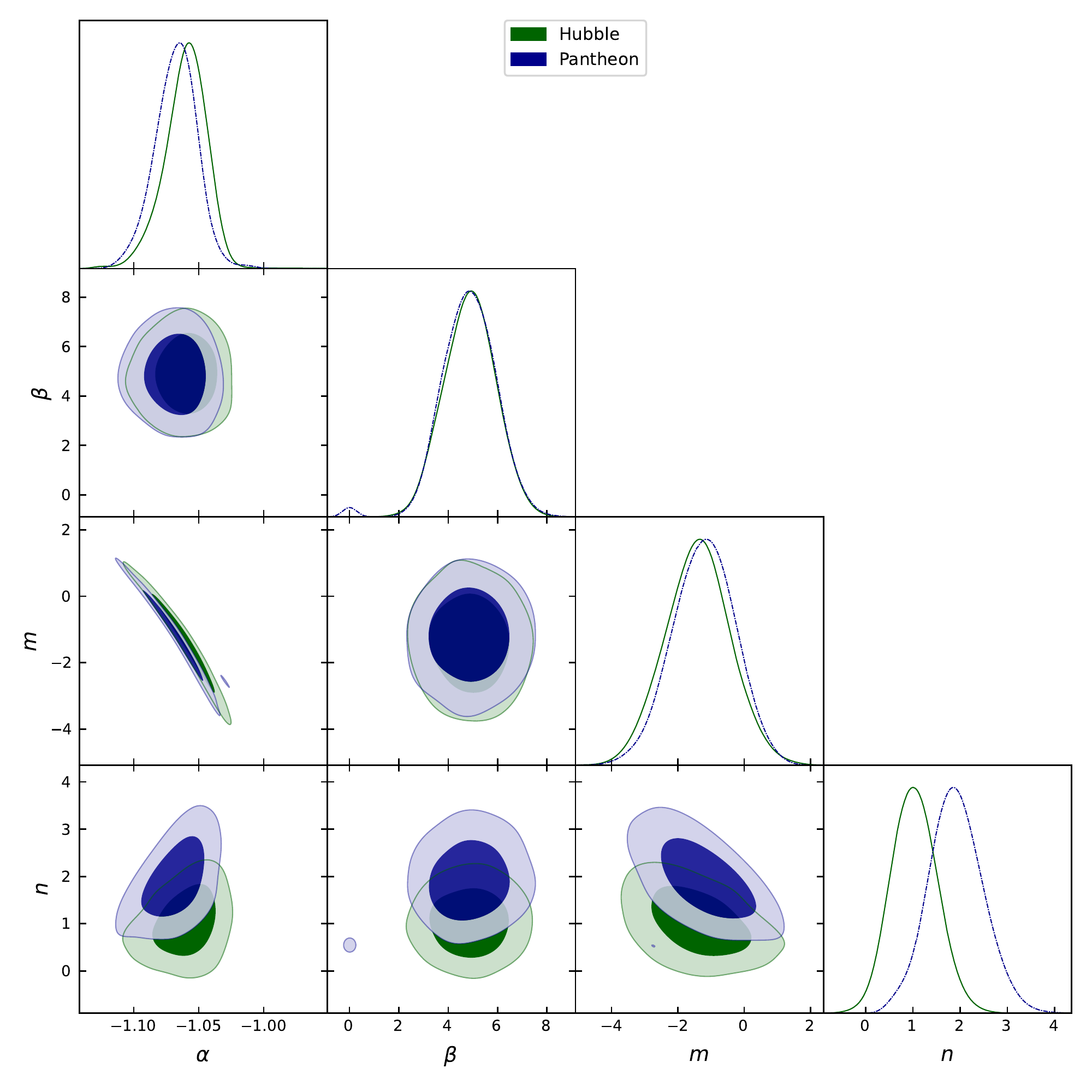}
\caption{Contour of the $1-\sigma$ and $2-\sigma$ confidence levels from the Hubble and Pantheon data for the model parameters $\alpha$, $\beta$, $m$, and $n$.}
\label{Combine}
\end{figure}

\begin{table}[H]
\begin{center}
  \caption{The marginalized constraining results on model parameters $\alpha$, $\beta$, $\omega_0$, and $\omega_a$ are shown by using the Hubble and Pantheon SNe Ia samples.}
    \label{Table}

\begin{tabular}{|l|c|c|c|c|}
\hline 
Dataset          & $\alpha$ & $\beta$ & $\omega_0 =m$ & $\omega_a =n$\\
\hline
Hubble           &  $-1.06^{+0.018}_{-0.013}$ & $4.9^{+1.0}_{-1.0}$ & $-1.4^{+0.98}_{-0.98}$ & $1.03^{+0.49}_{-0.49}$  \\ \hline
Pantheon            & $-1.067^{+0.017}_{-0.015}$  & $4.8^{+1.2}_{-1.2}$ & $-1.21^{+0.94}_{-0.94}$ & $1.91^{+0.58}_{-0.58}$   \\ \hline
Hz+Pantheon           & $-1.07078^{+0.00054}_{-0.00054}$ & $5.0^{+0.01}_{-0.01}$ & $-1.0006^{+0.0099}_{-0.0099}$ & $1.927^{+0.01}_{-0.01}$  \\ \hline
BAO+Hz+Pantheon  & $-1.07574^{+0.00049}_{-0.00049}$ & $5.0003^{+0.0098}_{-0.0098}$  & $-1.008^{+0.01}_{-0.01}$ & $1.91^{+0.01}_{-0.01}$ \\ \hline
\end{tabular}
\end{center}
\end{table}

\begin{figure}[]
\centering
\includegraphics[scale=0.65]{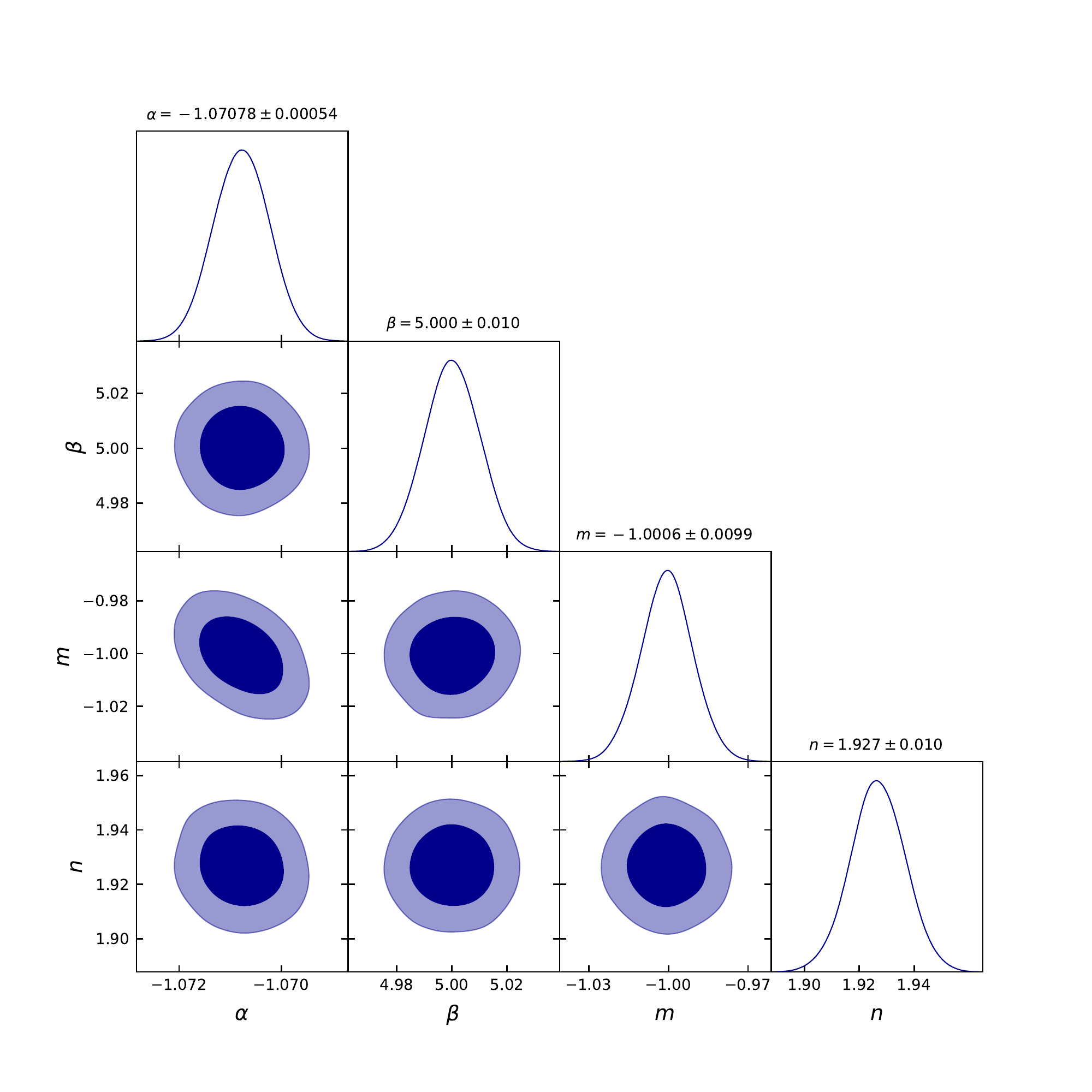}
\caption{Contour of the $1-\sigma$ and $2-\sigma$ confidence levels from the combination $Hz+Pantheon$ for the model parameters $\alpha$, $\beta$, $m$, and $n$.}
\label{Hz+Pan}
\end{figure}

\begin{figure}[]
\centering
\includegraphics[scale=0.65]{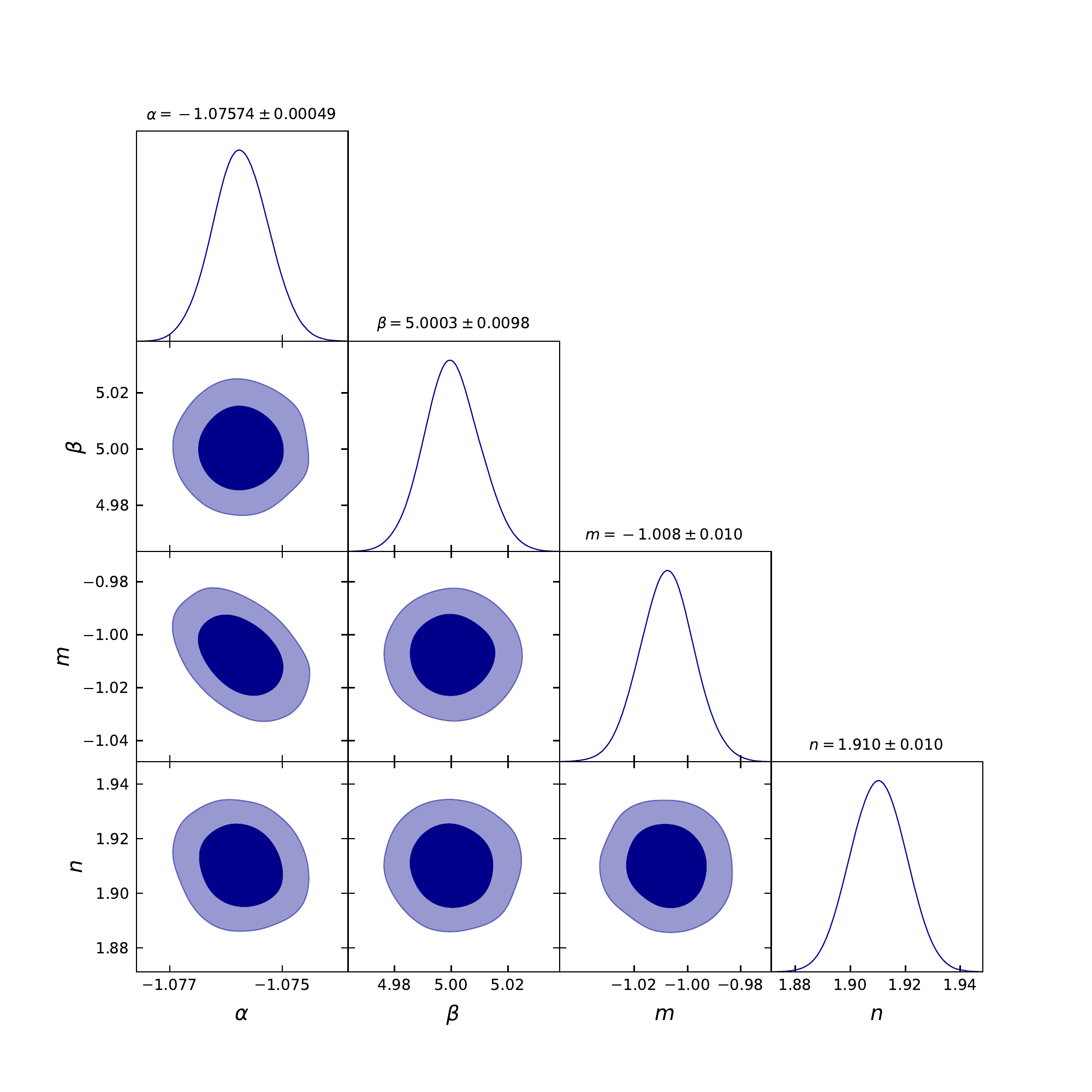}
\caption{Contour of the $1-\sigma$ and $2-\sigma$ confidence levels from the combination $BAO+Hz+Pantheon$ for the model parameters $\alpha$, $\beta$, $m$, and $n$.}
\label{BAO+Hz+Pan}
\end{figure}

\end{widetext}

\section{cosmological parameters}
\label{section 5}

The evolution of the density parameter, pressure, deceleration parameter, and the effective equation of state parameter is presented below.
\begin{figure}[]
\includegraphics[scale=0.6]{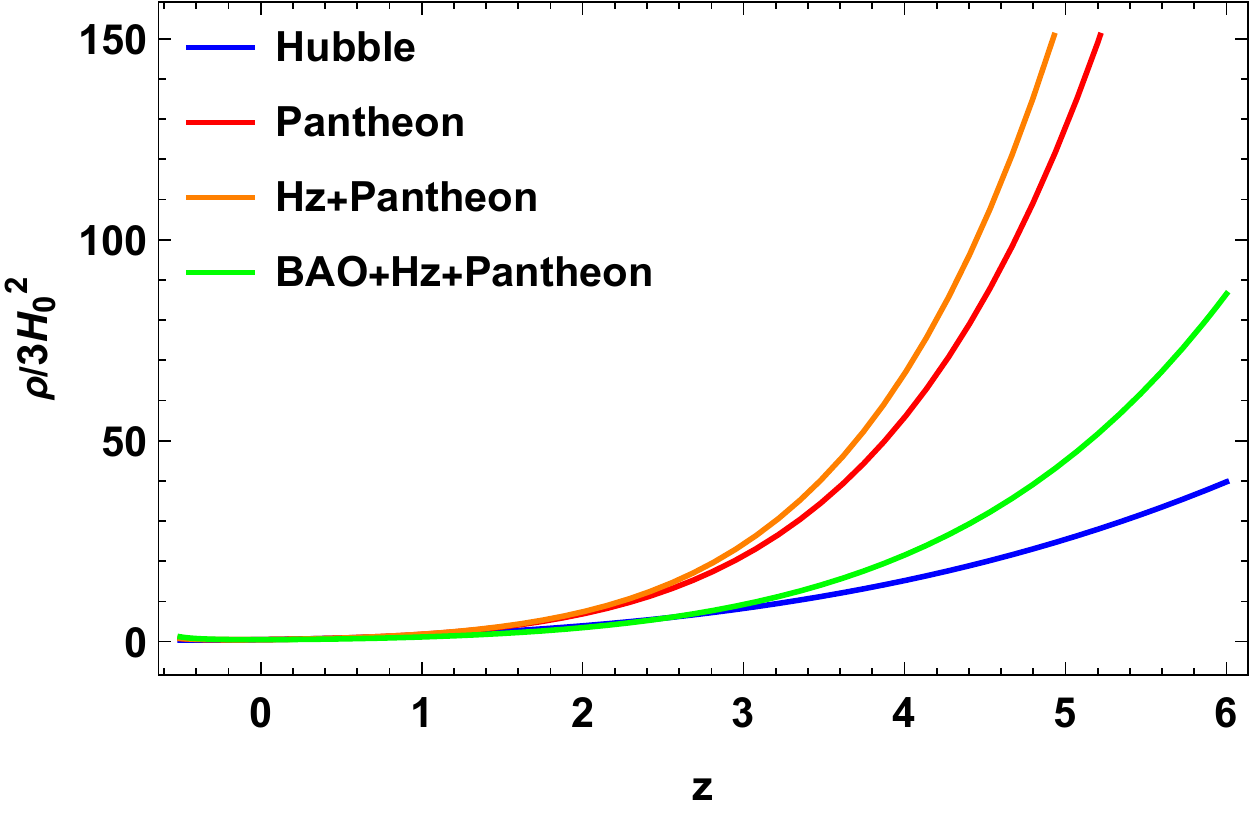}
\caption{The evolution of the density parameter $\rho$ vs redshift $z$ for the constrained values of model parameters.}
\label{density}
\end{figure}

\begin{figure}[]
\includegraphics[scale=0.6]{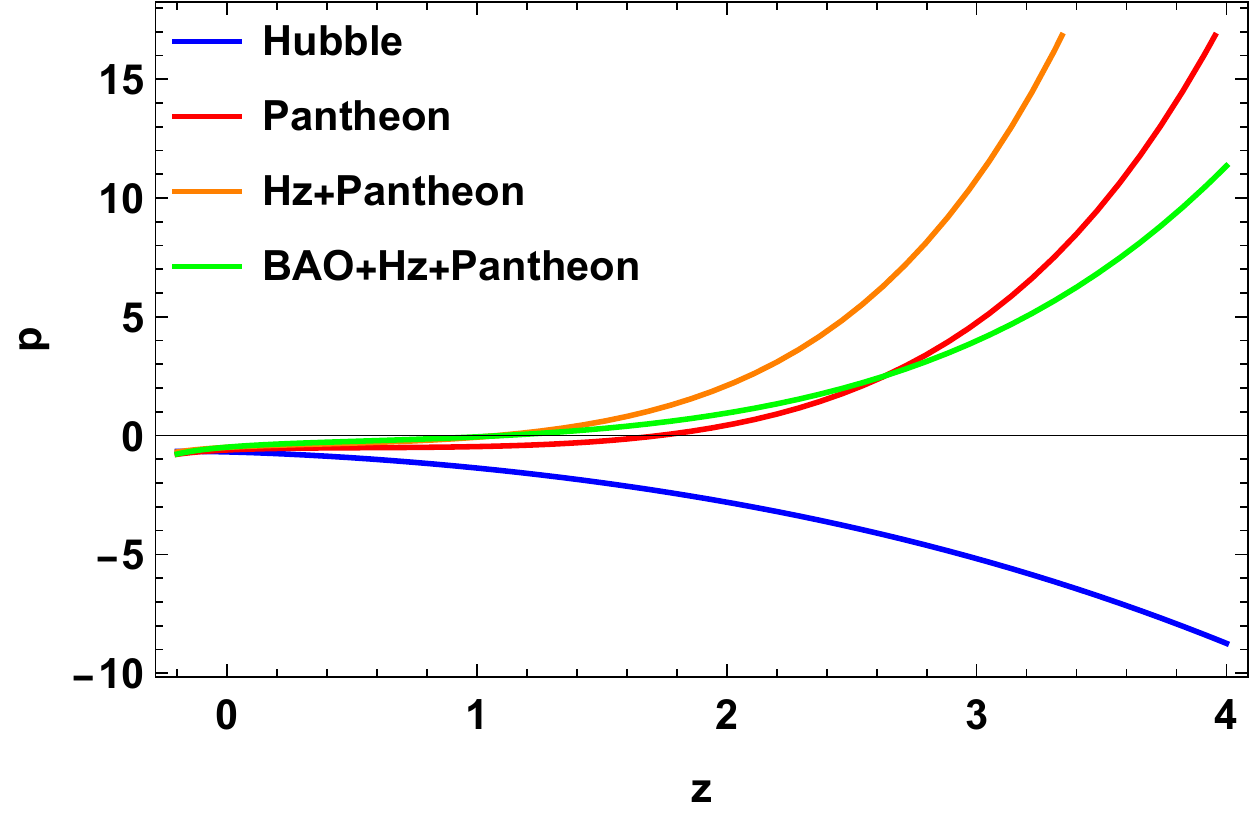}
\caption{The evolution of the pressure $p$ versus redshift $z$ for the constrained values of model parameters.}
\label{pressure}
\end{figure}
\begin{figure}[H]
\includegraphics[scale=0.6]{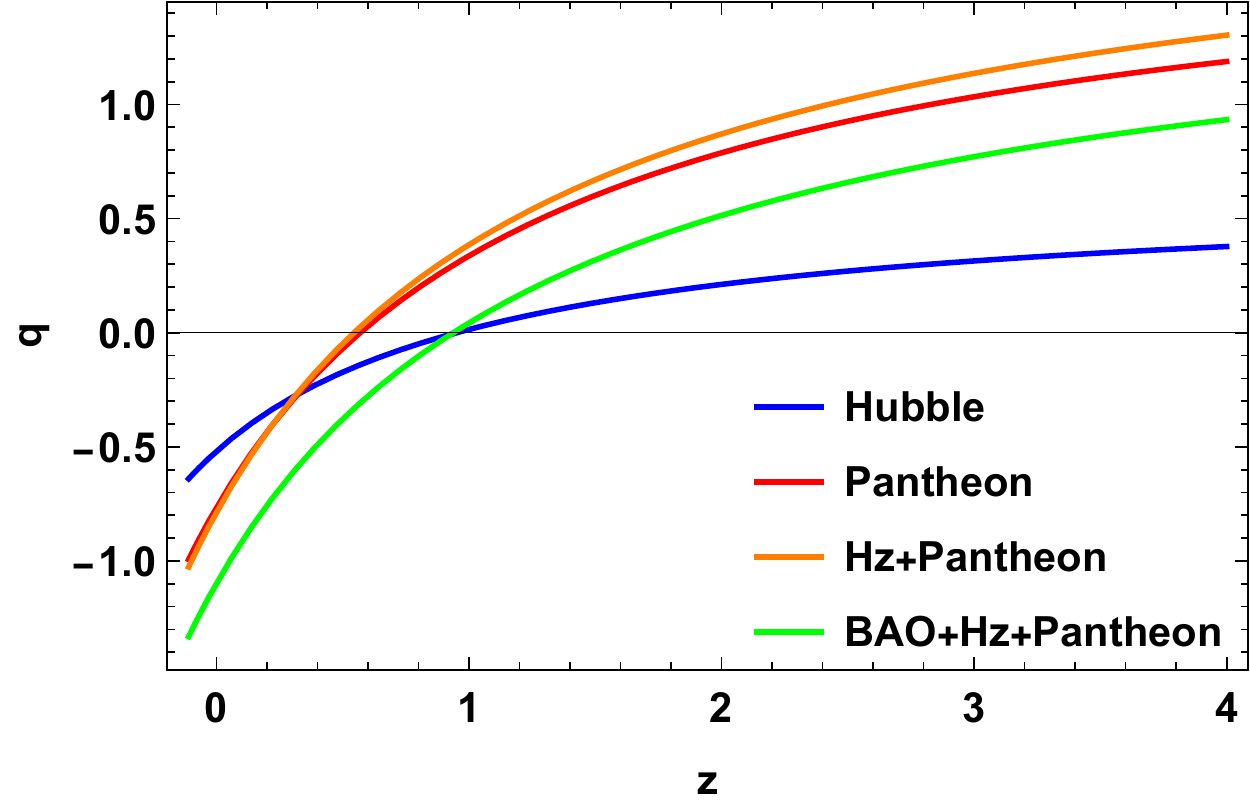}
\caption{The evolution of the deceleration parameter $q$ versus redshift $z$ for the constrained values of model parameters.}
\label{q}
\end{figure}

Clearly, figure \ref{density} shows that the positive behavior of energy density is as expected and decreases with the expansion of the universe in the present and far future. Figure \ref{pressure} depicts the negative behavior of the pressure $p$, indicating the late-time cosmic acceleration of the universe. It can be seen that the $Pantheon$, $Hz+Pantheon$ and $BAO+Hz+Pantheon$ data exhibit a different evolution of pressure in the past, but negative behavior at the present stage supports acceleration.\\
The deceleration parameter, a dimensionless representation of the second-order time derivative of the scale factor is presented in figure \ref{q}. For the proposed model, the deceleration parameter shows a signature flip ranging from 0.55 to 0.95, which is consistent with the analysis done in references \cite{Santos/2016,Garza/2019}. The present value of the deceleration parameter is $q_0=-0.52^{+0.6}_{-0.8}$, $q_0=-0.77^{+0.3}_{-0.3}$, $q_0=-0.78^{+0.05}_{-0.05}$ and $q_0=-1.1^{+0.08}_{-0.08}$ corresponding to the model parameters constrained by the $Hubble$, $Pantheon$, $Hz+Pantheon$ and $BAO+Hz+Pantheon$ datasets, respectively \cite{Sergio/2012,Cunha/2008,Camarena/2020}.  It clearly shows the late-time cosmic acceleration of the universe and the deceleration expansion in the past. 

\begin{figure}[H]
\includegraphics[scale=0.6]{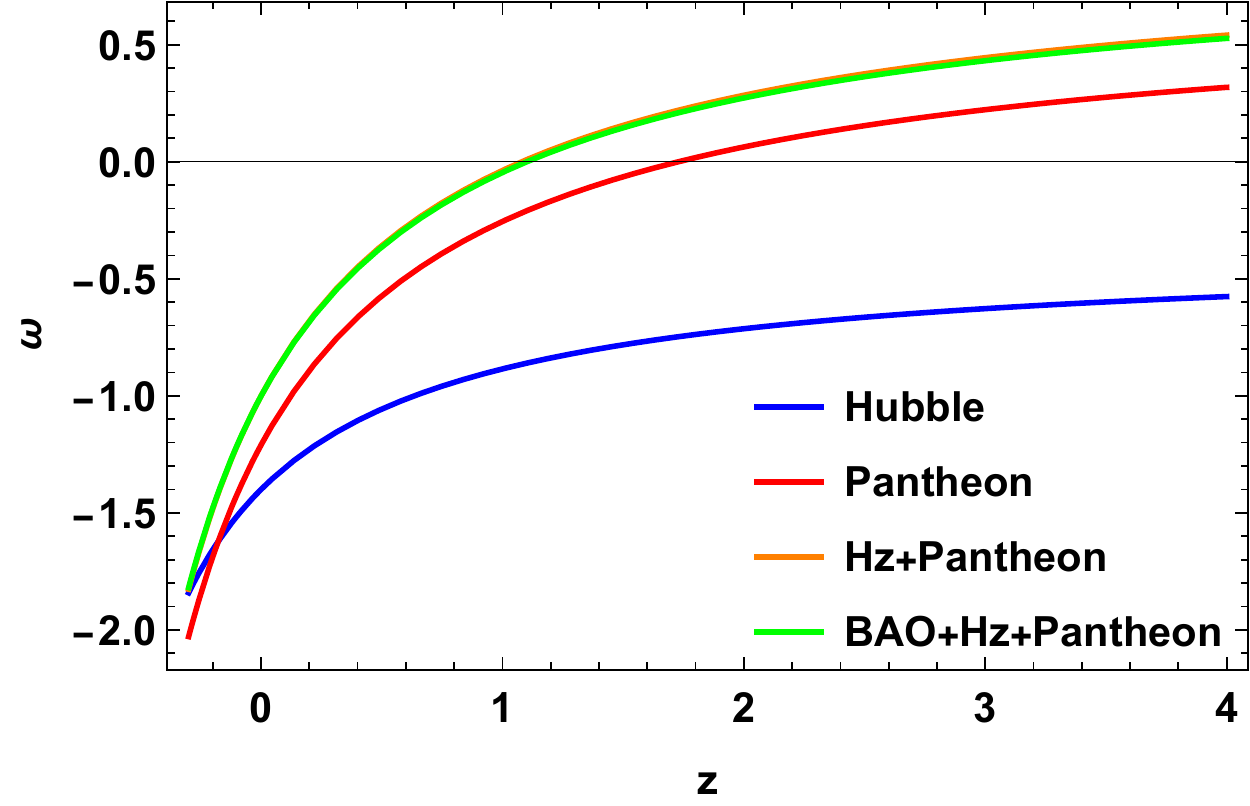}
\caption{The evolution of the effective equation of state parameter $\omega$ versus redshift $z$ for the constrained values of model parameters.}
\label{w}
\end{figure}

By fitting the model to the observational data, we find the present value of the EoS parameter corresponding to the constrained values of the model parameters as $\omega_{0}=-1.4^{+0.98}_{-0.98}$, $\omega_{0}=-1.21^{+0.94}_{-0.94}$, $\omega_{0}=-1.0006_{-0.0099}^{+0.0099}$ and $\omega_{0}=-1.008_{-0.01}^{+0.01}$, respectively \cite{Gong/2007,Novosyadlyj/2012,Suresh/2014,Mukherjee/2016}.
The evolution of $\omega$ is plotted in figure \ref{w}. We see that 
$\omega<0$ and crosses the -1 barrier in the redshift range $0.2-0.4$. 
We have demonstrated that the EoS parameter $\omega$ evolves from the quintessence region $(\omega>-1)$ to the phantom regime $(\omega<-1)$ \cite{Puxun/2010}.
At low redshift, this evolution of the dark energy parameter of EoS favors the various observational data sets \cite{Melchiorri/2003,Alam/2004}. It is fascinating that the phantom phase is twice as probable than the quintessence phase. Furthermore, the cosmological dynamics of the universe with such a phantom energy component possess many exciting features \cite{Caldwell/2003}. Detailed analyses of the Lagrangians describing phantom energy reveal that in some cases, the universe with phantom energy ends in a "big rip," whereas in others, it approaches the de Sitter expansion asymptotically \cite{McInnes}.
This is not a generic feature of the scenario but rather the consequence of the CPL parametrization.  It is clear that the CPL parameterization is in good agreement with $\Lambda$CDM model for $Pantheon$ samples. Another interesting point is the uncertainties associated with the value of $\omega$ vary according to the deviations from the $\Lambda$CDM. Another interesting thing is $H_0$ tension. Several attempts to resolve this tension with new physics have relied on extended cosmological models. According to the references \cite{Vagnozzi/2020,Valentino/2016}, a phantom-like component with an effective equation of state $\omega \approx -1.29$ can solve the current tension between the Planck data set and other priors in an extended $\Lambda$CDM scenario. We observed that our obtained model lies in the Phantom phase with the equation of state $-1.0006\leq \omega \leq -1.4$. As a result, our obtained model may be able to alleviate some tension at the present point.
 
\section{Conclusion}
\label{section 6}

This section will discuss the results obtained in the previous sections for the model developed in the Weyl-type $f(Q,T)$ gravity. The mystery of dark energy makes it highly implausible that the universe's expansion is actually accelerating. Although vacuum quantum energy can explain this dynamical effect via the cosmological constant in GR, the aforementioned significant and persistent problems with $\Lambda$ urge alternative explanations.\\

In the present work, we have considered a newly proposed Weyl-type $f(Q, T)$ gravity, as an alternative and effective modified theory of gravity. In the framework of the proper Weyl geometry, the scalar non-metricity $Q$ is wholly determined by the magnitude of the Weyl vector $w_{\lambda}$.  We used the linear combination of the nonmetricity $Q$ and the trace $T$ of the energy-momentum tensor, i.e., $f(Q, T)=\alpha Q+\frac{\beta}{6\kappa^2}T$, where $\alpha$ and $\beta$ are constants.\\ 

In section \ref{section 3}, we used the widely used Chevallier-Polarski-Linder (CPL) parametrization $\omega(z) =\omega_0+\omega_a \left( \frac{z}{1+z}\right) $, where $\omega_0$ and $\omega_a$ are constants to reconstruct the effective equation of state. The CPL parametrization is unaffected by any prior assumptions about the nature of dark energy. Further, we confronted the Hubble parameter with the latest observational datasets, namely the $Hubble$, $Pantheon$, and $BAO$ datasets to constrain the model parameters $\alpha$, $\beta$, $\omega_0=m$ and $\omega_a=n$ in table \ref{Table}. We have obtained the best-fit values of the model parameters and $1-\sigma$ and $2-\sigma$ confidence regions in figures. \ref{Combine}, \ref{Hz+Pan} and \ref{BAO+Hz+Pan}.\\

The probability of CPL parametrization in the non-phantom regime is less than in the phantom regime. This undeniably indicates the strong tendency of observational data sets toward dark energy crossing a phantom divide. It is noted that our model lies in a phantom phase with the equation of state $-1.0006 \leq \omega \leq -1.4$. The EoS parameter evolves from the quintessence region to the phantom regime. It is clear that the constructed model allows for a wide range of $\omega$ values.  In the course of the universe's evolution, the deceleration parameter causes a dynamic change from deceleration to acceleration. Moreover, we obtained a deceleration parameter in case of $BAO+Hz+Pantheon$ as $q_0=-1.1^{+0.08}_{-0.08}$ which deviates from $\Lambda$CDM at $1-\sigma$ level.  As a result, the phantom cosmology may be capable of composing the underlying mechanism for dark energy. According to the references \cite{Vagnozzi/2020,Valentino/2016}, a phantom-like component with an effective equation of state $\omega \approx -1.29$ can solve the current tension between the Planck data set and other priors in an extended $\Lambda$CDM scenario. We observed that our obtained model of $f(Q,T)$ also lies in the phantom phase with the equation of state $-1.0006\leq \omega \leq -1.4$. As a result, our obtained model may be able to alleviate some tension at the present point. The difference between the values and behaviors of the deceleration and equation of state parameters from the $\Lambda$CDM model points to a new dark energy alternative. The $f(Q,T)$ theory could provide a promising explanation for the accelerated expansion of the universe and new cosmic findings.

\section*{Data Availability Statement}

There are no new data associated with this article.

\section*{Acknowledgements}

GG acknowledges University Grants Commission (UGC), New Delhi, India for awarding Junior Research Fellowship (UGC-Ref. No.: 201610122060). SA acknowledges BITS-PIlani, Hyderabad Campus for Institute Fellowship. PKS acknowledges IUCAA, Pune, India for providing support through the visiting Associateship program. We are very much grateful to the honorable referees and to the editor for the illuminating suggestions that have significantly improved our work in terms of research quality, and presentation.

\end{document}